\documentclass[10pt,letterpaper]{article}
\usepackage[top=0.85in,left=2.75in,footskip=0.75in]{geometry}

\usepackage{amsmath,amssymb}

\usepackage{changepage}

\usepackage[utf8x]{inputenc}
\usepackage[greek,english]{babel}

\usepackage{textcomp,marvosym}

\usepackage{cite}

\usepackage{nameref,hyperref}

\usepackage[right]{lineno}

\usepackage{microtype}
\DisableLigatures[f]{encoding = *, family = * }

\usepackage[table]{xcolor}

\usepackage{array}

\newcolumntype{+}{!{\vrule width 2pt}}

\newlength\savedwidth

\newcommand\thickhline{\noalign{\global\savedwidth\arrayrulewidth\global\arrayrulewidth 2pt}%
\hline
\noalign{\global\arrayrulewidth\savedwidth}}

\raggedright
\usepackage[aboveskip=1pt,labelfont=bf,labelsep=period,justification=raggedright,singlelinecheck=off]{caption}

\makeatletter
\renewcommand{\@biblabel}[1]{\quad#1.}
\makeatother

\usepackage{lastpage,fancyhdr,graphicx}
\usepackage{epstopdf}
\fancyheadoffset[L]{2.25in}
\fancyfootoffset[L]{2.25in}
\begin{document}
\vspace*{0.2in}

\begin{flushleft}
{\Large
\textbf\newline{DNA coding and Gödel numbering} 
}
\newline
\\
Argyris Nicolaidis\textsuperscript{1},
Fotis Psomopoulos\textsuperscript{2,3*}
\\
\bigskip
\textbf{1} Theoretical Physics Department, Aristotle University of Thessaloniki, Greece \\
\textbf{2} Institute of Applied Biosciences, Centre for Research and Technology Hellas, Thessaloniki, Greece\\
\textbf{3} Department of Molecular Medicine and Surgery, Karolinska Institutet, Stockholm, Sweden\\
\bigskip

%
%

* corresponding authors: nicolaid@auth.gr, fpsom@certh.gr

\end{flushleft}
\section*{Abstract}
Evolution consists of distinct stages: cosmological, biological, linguistic. Since biology verges on natural sciences and linguistics, we expect that it shares structures and features from both forms of knowledge. Indeed, in DNA we encounter the biological "atoms", the four nucleotide molecules. At the same time these four nucleotides may be considered as the "letters" of an alphabet. These four "letters", through a genetic code, generate biological "words", "phrases", "sentences" (aminoacids, proteins, cells, living organisms).

In this spirit we may consider equally well a DNA strand as a mathematical statement. Inspired by the work of Kurt Gödel, we attach to each DNA strand a Gödel's number, a product of prime numbers raised to appropriate powers. To each DNA chain corresponds a single Gödel's number $G$, and inversely given a Gödel's number $G$, we can specify the DNA chain it stands for. Next, considering a single DNA strand composed of $N$ bases, we study the statistical distribution of $g$, the logarithm of $G$. Our assumption is that the choice of the $m$th term is random and with equal probability for the four possible outcomes. The "experiment", to some extent, appears as throwing $N$ times a four-faces die. Through the moment generating function we obtain the discrete and then the continuum distribution of $g$. There is an excellent agreement between our formalism and simulated data. At the end we compare our formalism to actual data, to specify the presence of traces of non-random dynamics.


\section*{Introduction}

Everything is under the realm of evolution. It all started with what is traditionally known in cosmology as the "big bang". The universe originated from a state of high temperature and high density (13.7 billion years ago). The subsequent expansion gave rise to a cool universe and led to the formation of galaxies and stars ~\cite{weinberg1972gravitation01, Mukhanov2005sc02, 2003itcRyden03}. The cosmological stage of evolution was followed by the biological stage. In a "friendly" planet, our earth, appeared the first cell (somehow 3.5 billion years). The biological evolution was rapid, as it was studied and advocated by Darwin. Furthermore, the Darwinian point of view was supported by the discovery of the building blocks of the biological organisms, the DNA ~\cite{Dennett1996DENDDI04, dawkins-selfish-gene-2006-05}. The third evolutionary stage involved the development of human language. Language allowed an effective communication among the members of a human group, helped in transferring information from one generation to another, creating an endless semiotic process ~\cite{chomsky06, Nowak07, Nicolaidis_2009_08}. It seems that presently we are in the eve of the fourth stage of evolution, where through technology and artificial intelligence we are led to posthuman, a state beyond the human.

All systems we know, cosmos and the physical world, biological systems, linguistic systems are highly generative systems. Few constituent elements form larger blocks, then these blocks following basic rules form a huge variety of entities. This limitlessness of a highly generative system has been described as "making infinite use of finite means" ~\cite{Hauser09}.

In actual terms we may notice that nature, despite its immense variety can be analyzed and understood as a collection of few building blocks, the elementary particles (quarks, leptons, gauge particles). The elementary particles interact and form (or transformed to) larger compounds (nuclei, molecules, galaxies) via the four well known interactions. We may view the elementary particles as constituting an "alphabet", and the interactions as providing the "rules of composition" (or "grammatical rules") to create the larger configurations (texts). Within this analogy scheme, it is rather significant that the ancient Greeks were using the same word (\textgreek{στoιχεία}) to denote both the letters of the alphabet and the constitutive elements of the universe.

Biology uses another exemplary generative system. In DNA we encounter the biological "atoms", the four nucleotide molecules (adenine, guanine, cytosine, thymine). These four nucleotides get composed to form larger structures, the DNA sequences, amino acids, proteins, living organisms. From another point of view the four nucleotides may be considered as not simply the constituents of biological structures, but as the "letters" of a language. These "letters" give rise to biological "words", "phrases", "sentences". The biological "words" or "phrases" act like signs, receiving – registering – transferring information, executing specific functions, favoring or disfavoring a biological process.  It is an open and a highly interesting question if the biological "text" follows an internal logic, or a syntax. Noam Chomsky, who revolutionized linguistic research, emphasized that the human faculty of language appears to be organized like the biological genetic code - hierarchical, generative, recursive, and virtually limitless with respect to its scope of expression. 

Having this in mind, we may consider that for a single DNA strand (a chain of bases, where two consecutive bases are bound together by a covalent bond) each of the four bases stands for a letter, establishing a 4-letter alphabet. A succession of these letters in a DNA strand may represent a word or a phrase in a biological language. Equally well, each base may represent a symbol in an axiomatic system. The most formalized language is mathematics and a succession of these symbols may correspond to a mathematical theorem. In this spirit, we are entitled to follow the work of Kurt Gödel ~\cite{godel10}. Gödel made an immense impact upon scientific and philosophical thinking in the 20th century, by establishing the incompleteness theorem. To prove this theorem, Gödel developed a technique now known as Gödel numbering, which codes formal expressions by natural numbers. This numbering was adopted and allowed to codify a DNA strand by a Gödel’s number $G$, a product of prime numbers raised to appropriate powers ~\cite{Nicolaidis_2016_11}. Inversely, given a Gödel’s number $G$ we are able to specify the DNA strand it stands for.

In the present work we consider a single DNA strand composed of $N$ bases and we study the statistical distribution of $g$, the logarithm of $G$. Our assumption is that the choice of the $n$th term ($n = 1, 2, 3, \ldots, N-1, N$) is random and with equal probability for the four possible outcomes. The "experiment", to some extent, appears as throwing $N$ times a four-faces die. Through the moment generating function we obtain the discrete and then the continuum distribution of $g$. There is an excellent agreement between our formalism and simulated data. At the end we compare our formalism to actual data, to specify the presence of traces of non-random dynamics. The presence of non-randomness is considered as a sign of information processing.

\section*{Materials and methods}

\subsection*{Kurt Gödel and DNA Coding}

It is well known that every integer larger than 1 can be written as a product of prime numbers. This factorization is unique and primes can be considered as the "basic building blocks" of the natural numbers.

Kurt Gödel suggested a numbering, assigning a unique natural number to each formula appearing within a mathematical theory or a formal language. He used a system based on prime factorization. First a positive number was given to each basic symbol in the formal language. To encode an entire formula, which is a sequence of $N$ symbols, Gödel considered the product of the first $N$ primes. Each prime was raised to an appropriate power. The $m$th prime was raised to the power corresponding to the positive number for the symbol appearing in the $m$th place ~\cite{godel10}. The Gödel number $G$ thus obtained is unique. Inversely, given a Gödel’s number $G$, we can decode it and find the formula it represents.

In our case we are dealing with the DNA, which is composed of four "basic building blocks": adenine (A), cytosine (C), guanine (G) and thymine (T). Out of these four letters, A, C, G, T, we may form a DNA strand, to be considered as a word or a phrase in a biological language. We start by assigning a number to each of the four letters. Clearly the choice is not unique. In our case (Eq~(\ref{eq:aminoacidAssignment})) we may choose:

\begin{eqnarray}
\label{eq:aminoacidAssignment}
\displaystyle
    \begin{array}{lcr}
    \mbox{\#(A)} & = & 1 \\
    \mbox{\#(C)} & = & 2 \\
    \mbox{\#(G)} & = & 3 \\
    \mbox{\#(T)} & = & 4 \end{array}\
\end{eqnarray}

For a succession of $N$ DNA bases, we pick up the first $N$ primes and raise each of them to the corresponding power ~\cite{Nicolaidis_2016_11}. For example, for the DNA strand $ATCG$ the corresponding Gödel number $G$ is:

\begin{eqnarray}
\label{eq:exampleSeqToGodel}
\displaystyle
G(\texttt{ATCG})=2^1 \cdot 3^4 \cdot 5^2 \cdot 7^3 = 1389150
\end{eqnarray}

Given a Gödel number $G$, we can decode it by prime factorization and find the DNA strand it represents ~\cite{Nicolaidis_2016_11}. For example, consider the number $G = 9450$. By factorizing $G$ as a product of the prime numbers $2, 3, 5, 7, 11, \ldots$ we may find out how many $2, 3, 5, 7, 11, \ldots$ are hidden in the number. In our example:

\begin{eqnarray}
\label{eq:exampleGodelToSeq}
\displaystyle
G = 9450 = 2^3 \cdot 3^4 \cdot 5^1
\end{eqnarray}

and therefore, the above $G$ number stands for the DNA sequence \texttt{GTA}.

Gödel’s numbering allows to obtain a quantitative measure of the difference among the various DNA strands. Considering for example a reference DNA strand (represented by $G_1$) and another strand (represented by $G_2$), we define the difference $\Delta$ between the reference stand and the second strand by:

\begin{eqnarray}
\label{eq:strandDifferenceGodel}
\displaystyle
\Delta = \prod_{j} \frac{1}{p_j}
\end{eqnarray}

where $p_j$ stands for a prime umber where the DNA bases differ in the corresponding $j$th place ~\cite{Nicolaidis_2016_11}. 

It is interesting to study the distribution of the Gödel’s number $G$ over the positive integers. Our working assumptions are the following:

\begin{enumerate}
	\item{The choice of the DNA base at the nth step is independent of the previous or the next choices}
	\item{The choice is random and with equal probability for the four possible outcomes (A, C, G, T)}
\end{enumerate}

Clearly these assumptions may be relaxed in a more detailed study. 

In general, for a DNA chain with $N$ bases:

\begin{eqnarray}
\label{eq:calculateGodelNumber}
\displaystyle
G = p_1^{a_1} \cdot p_2^{a_2} \cdot p_3^{a_3} \cdot \ldots \cdot p_N^{a_N}	
\end{eqnarray}

It appears more appropriate to work with the logarithm of $G$, $g = log(G)$.

\begin{eqnarray}
\label{eq:calculateLogGodelNumber}
\displaystyle
\begin{split}
g &= g_1 + g_2 + g_3 + \ldots + g_N \\
 &= a_1 \cdot \log(p_1) + a_2 \cdot \log(p_2) + a_3 \cdot \log(p_3) + \ldots + a_N \cdot \log(p_N) \\
\end{split}
\end{eqnarray}

where $p_1, p_2, p_3, \ldots, p_N$ are the successive prime numbers ($2, 3, 5, \ldots$), while $a_n$ is the random variable denoting the outcome in the $n$th position of the DNA sequence. 

The variable $g$ appears as the sum of the numbers we get when we roll a fair die $N$ times. However, there are two differences with an ordinary die:

\begin{enumerate}
	\item{Our die has four faces (1, 2, 3 and 4) rather than six}
	\item{Most importantly, each time we roll the die, the outcome is measured with a different scale. The first roll is scaled with $\ln(p_1)$, the second with $\ln(p_2)$, $\ldots$, the $N$th roll with $\ln(p_N)$.}
\end{enumerate}

To proceed further we study the moment-generating function, defined as the expectation of the random variable $e^{t \cdot g}$.

\begin{eqnarray}
\label{eq:calculateExpectionRandomVariable}
\displaystyle
M_g(t) = E[e^{t \cdot g}]	
\end{eqnarray}

Taking into account our working assumptions we obtain:

\begin{eqnarray}
\label{eq:calculateExpectionRandomVariableGodel}
\displaystyle
\begin{split}
M_g(t) &= \prod_{n = 1}^{N} E[e^{t \cdot g_n}] =  \\
 &= \prod_{n = 1}^{N} \Big[ \frac{1}{4} \cdot \Big( e^{t \cdot \log(p_n)} + e^{2 \cdot t \cdot \log(p_n)} + \ldots + e^{3 \cdot t \cdot \log(p_n)} + e^{4 \cdot t \cdot \log(p_n)} \Big) \Big]
\end{split}
\end{eqnarray}

A Taylor expansion in $t$ provides:

\begin{eqnarray}
\label{eq:taylorExpansion}
\displaystyle
M_g(t) = \prod_{n = 1}^{N} \Big[ 1 + t \cdot \overline{g_n} + \frac{1}{2} \cdot t^2 \cdot  \overline{g^2_n} + \ldots \Big]	
\end{eqnarray}

By keeping the linear term in $t$ we obtain the mean value $\overline{g}$:

\begin{eqnarray}
\label{eq:taylorExpansionMeanValue}
\displaystyle
\overline{g} = \sum_{n = 1}^{N} \overline{g_n} = 2.5 \cdot \sum_{n = 1}^{N} \log{p_n}
\end{eqnarray}

The expectation value of $g^2$, denoted by $\overline{g^2}$, is found to be:

\begin{eqnarray}
\label{eq:taylorExpansionExpectationValue}
\displaystyle
\overline{g^2} = \sum_{n = 1}^{N} \overline{g^2_n} + \sum_{i \neq j} 2 \cdot \overline{g_i} \cdot \overline{g_j}
\end{eqnarray}

The variance $var(g)$ is obtained by:

\begin{eqnarray}
\label{eq:varianceG}
\displaystyle
var(g) = \overline{g^2_n} - (\overline{g_n})^2 = 7.5 \cdot \sum_{n = 1}^{N} \log^2p_n - (2.5)^2 \cdot \sum_{n = 1}^{N} \log^2p_n = 1.25 \cdot \sum_{n = 1}^{N} \log^2p_n
\end{eqnarray}

From the discrete distribution we may move to a continuum distribution, considering a normal distribution:

\begin{eqnarray}
\label{eq:continuousDistribution}
\displaystyle
P(g) = \frac{1}{\sqrt{2 \cdot \pi \cdot \sigma^2}} \cdot \exp \Big( - \frac{(g - \overline{g})^2}{2 \cdot \sigma^2} \Big)
\end{eqnarray}

where $\sigma^2 = var(g)$.

In our case, and defining:

\begin{eqnarray}
\label{eq:defineP1}
\displaystyle
P_1 \equiv \sum_{n = 1}^{N} \log p_n
\end{eqnarray}

\begin{eqnarray}
\label{eq:defineP2}
\displaystyle
P_2 \equiv \sum_{n = 1}^{N} \log^2 p_n
\end{eqnarray}

we obtain

\begin{eqnarray}
\label{eq:continuousDistributionWithP}
\displaystyle
P(g) = \frac{1}{\sqrt{2 \cdot \pi \cdot 1.25 \cdot P_2}} \cdot \exp \Big( - \frac{(g - 2.5 \cdot P_1)^2}{2 \cdot 1.25 \cdot P_2} \Big)
\end{eqnarray}

\subsection*{Implementation}

\begin{figure}[!h]
\includegraphics[scale=0.5]{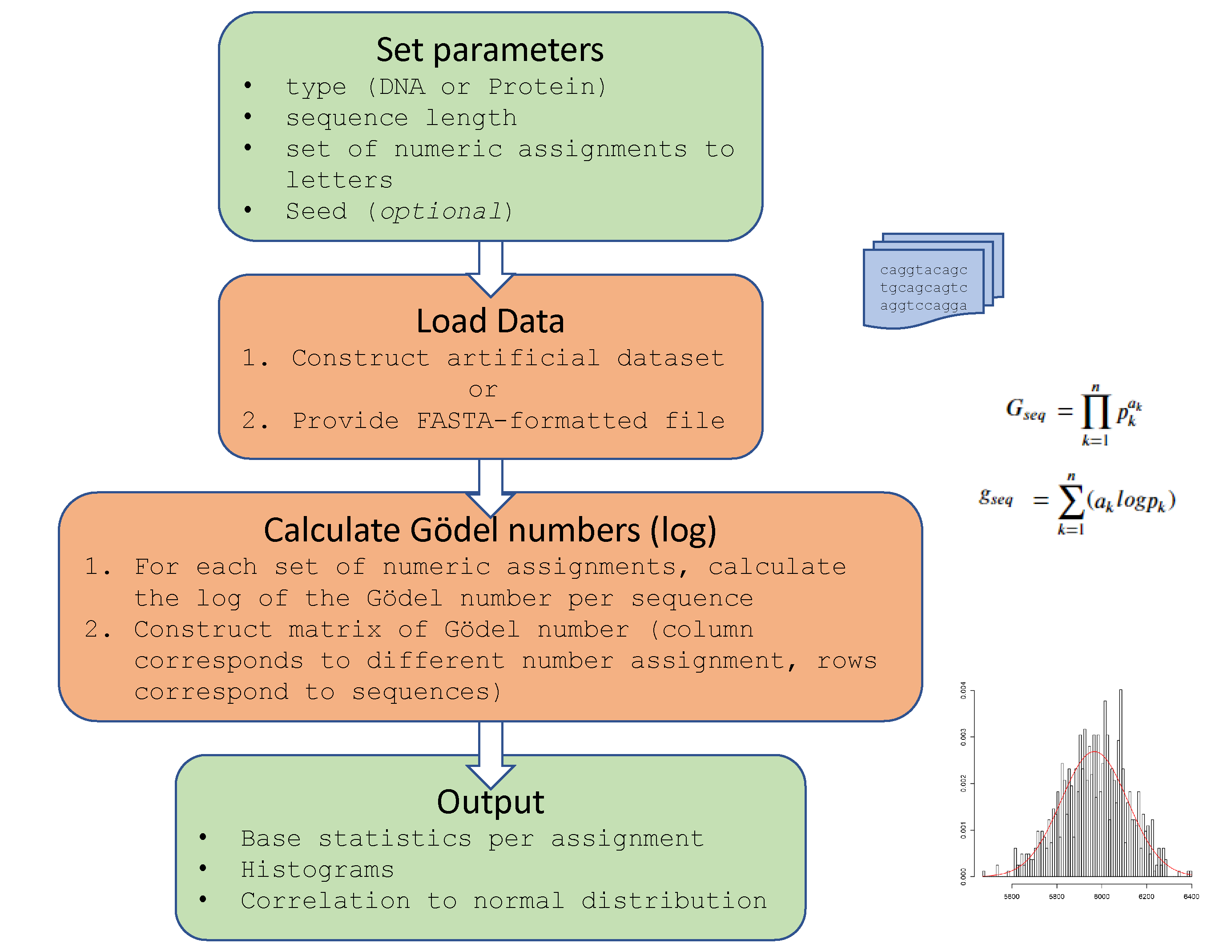}
\caption{{\bf General flow diagram.}
General flow diagram of the individual steps of the process.}
\label{figFlow}
\end{figure}

The overall method was implemented in R in the form of a Jupyter notebook ~\cite{Jupyter_12} (code is available in a \href{https://github.com/fpsom/godel}{GitHub repository} and also directly executable through the mybinder service ~\cite{project_jupyter-proc-scipy-2018_13}). An overview of the individual steps that take place in the notebook are shown in Fig~\ref{figFlow}.

In addition to the core of the notebook which implements the method as described earlier, we also provide the following functions:

\begin{itemize}
	\item{\texttt{\textbf{sieve}($n$)}: This function generates the list of prime numbers required for the calculation of the Gödel numbers that are identified under the limit n through the "sieve of Eratosthenes" ~\cite{eratosthenes_14} method.}
	\item{\texttt{\textbf{seqList}($x$, $type$)}: This function reads a FASTA formatted text file $x$ ($type$ parameter).}
	\item{\texttt{\textbf{createRandomSequencesBasedOnDistr}($count$, $length$, $prob=c(0.25,0.25,0.25,0.25)$, $fileNameRandSeqs$)}: This function produces count nucleotide sequences of length $length$, with the probability of appearance for each nucleotide adhering to the parameter prob, and stores them in fasta format in the file $fileNameRandSeqs$.}
	\item{\texttt{\textbf{assignSets}($randSequenceValues$, $type$)}: This function assigns the random values $randSequenceValues$ to each of the letters (based on the $type$ value). In each case, the exact random sequence is printed out, in order to retrieve the exact allocation if necessary.}
	\item{\texttt{\textbf{createRandomSequenceValues}($seedList$, $type$)}: This function creates a random permutation of values (i.e. 1 to 4, based on the $type$ value), for as many times as the number of seeds provided in the $seedList$.}
	\item{\texttt{\textbf{godelStatistics}($x$)}: This function prints out the summary statistics of the input variable $x$.}	
\end{itemize}

Finally, before running the core method, the following parameters must be set:
\begin{enumerate}
	\item{Define the ceiling of the primes list (e.g. 20000)}
    \item{Define whether full logging is required (TRUE/FALSE)}
    \item{Define how many different permutations of the letter – number assignments should be tried (e.g. 4 for nucleotide sequences)}
    \item{Define whether you need to specify a known seed to replicate (TRUE / FALSE)}
    \item{Define the type of sequence (DNA / AA)}
    \item{What is the sequence length (e.g. 361)}
    \item{What is the number of sequences to be processed (e.g. 821)}
\end{enumerate}

\section*{Results}

In order to evaluate the validity of our approach, we apply the method to two different datasets; the first is an artificial set of nucleotide sequences with pre-set values for the probabilities of appearance of each nucleotide. The second is a real-world dataset of 821 nucleotide sequences. In both cases, the assignment of numbers to nucleotides is the same as denoted in Eq~(\ref{eq:aminoacidAssignment}).

\subsection*{Simulated dataset}

\begin{figure}[!h]
\includegraphics[scale=0.4]{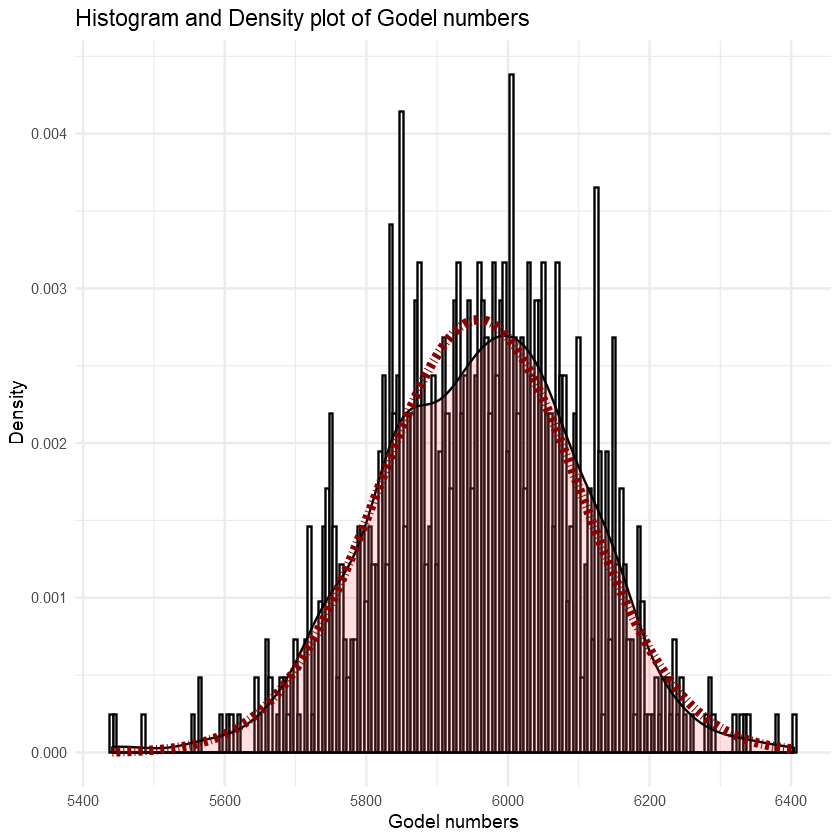}
\caption{{\bf Gödel numbers for artificial dataset.}
Distribution of Gödel numbers for the artificially generated nucleotide sequences. The corresponding theoretical curve is denoted by the red dotted line.}
\label{figArtificalHist}
\end{figure}

The simulated dataset is constructed through the use of the \texttt{createRandomSequencesBasedOnDistr} function. Setting the default distribution of nucleotide appearance in the sequences (0.25 for each letter), we expect the corresponding Gödel numbers to exhibit a normal distribution. Indeed, as evident in Fig~\ref{figArtificalHist} and listed in Table~\ref{table1}, the distribution closely follows the normal:

\begin{table}[!ht]
\centering
\caption{
{\bf Theoretical versus experimental for artificial dataset}}
\begin{tabular}{|l|l|l|l|}
\hline
  & Mean & std\\ \thickhline
Theoretical & 5966.88 & 142.75\\ \thickhline
Calculated & 5973.605 & 138.19 \\ \hline
\end{tabular}
\begin{flushleft} We observe that there is a very small (almost negligible) difference for both mean and the standard deviation values of the Gödel numbers. This is expected, given that the artificially-created sequences were following the expected distribution of nucleotides.
\end{flushleft}
\label{table1}
\end{table}

\subsection*{Real-world scenario}

We also attempted to evaluate the method using a real-world dataset of same-length nucleotide sequences. Specifically, we used 821 Ig/TCR nucleotide sequences of 361 length that were retrieved from a Chronic Lymphocytic Leukemia dataset.

\begin{figure}[!h]
\includegraphics[scale=0.4]{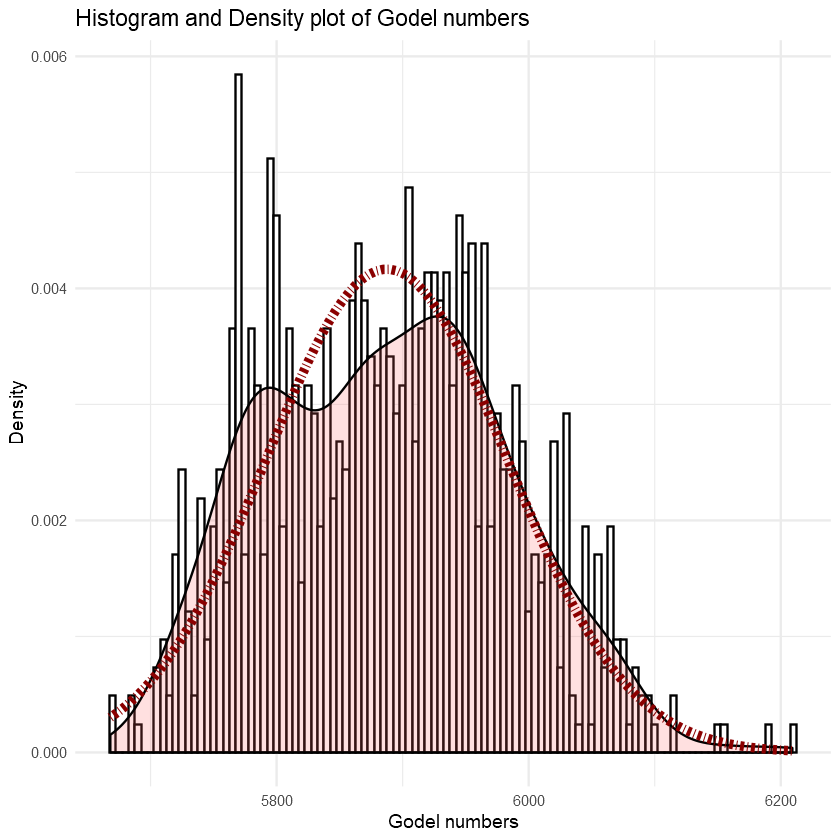}
\caption{{\bf Gödel numbers for real-world dataset.}
Distribution of Gödel numbers for the real-world nucleotide sequences. The corresponding theoretical curve is denoted by the red dotted line.}
\label{figRealWorldHist}
\end{figure}

As evident in  Fig~\ref{figRealWorldHist} and listed in Table~\ref{table2}, the distribution resembles a normal distribution; however, a close look at the particular metrics reveal that there is slight deviation in both mean and std (please note that the theoretical values remain the same, as we opted to use the exact same number of sequences and total length in both datasets).

\begin{table}[!ht]
\centering
\caption{
{\bf Theoretical versus experimental for real-world dataset}}
\begin{tabular}{|l|l|l|l|}
\hline
  & Mean & std\\ \thickhline
Theoretical & 5966.88 & 142.75\\ \thickhline
Calculated & 6199.06 & 106.69 \\ \hline
\end{tabular}
\begin{flushleft} We observe that there is a notable difference between the values obtained for the mean and standard deviation as they arise from the the theoretical and real data.
\end{flushleft}
\label{table2}
\end{table}

\section*{Discussion}

We established a correspondence between DNA and integer numbers. To each DNA strand we correspond a specific Gödel’s number. This is a single-value correspondence. Thus any possible DNA correlation is reflected as a correlation among numbers. Clearly this is a novel and important step in decoding the DNA structure.

As a first paradigm in our approach we considered the case of random DNA with equal probability for the four DNA bases. Employing well established techniques of statistical mechanics we obtained the probability distribution for the Gödel’s numbers. The agreement with simulated data is perfect, implying that our approach is meaningful.

Next we compared our formalism with "real data", namely the Ig/TCR nucleotide sequences. In this case, we see that a disagreement emerges. This disagreement may be attributed to different causes:

\begin{itemize}
	\item{The DNA strand is still randomly created, but with unequal probabilities for the four bases A, C, G. T. We plan to examine this possibility by considering the individual distributions for A, C, G, T.}
	\item{The frequent presence of biological "words". We  will check the possibility that specific bases trigger the appearance of other bases, or that the actual "words" are the aminoacids.}
	\item{The possibility that periodic patterns within DNA exist \cite{Kolias_Arxiv_2019}.}
	\item{DNA correlations arising from quantum entanglement \cite{QuantDNA_Arxiv_2011} \cite{Vedral_2018}.}
	\item{We established a correspondence between DNA and prime numbers. Clearly this coding is sensitive to the frequency appearance of the primes. This frequency is ultimately  connected to the Riemann’s hypothesis \cite{Sautoy2012}.}
	\item{The sample we used is not representative of the biological data.}
	\item{The complexity of a pre-assumed biological "language". We might be missing important aspects of the biological language.}
\end{itemize}

Finally, a significant constraint of this method is the requirement of "same-length" sequences. Although this is not generally applicable to meaningful biological sequences (i.e. genes and/or proteins), as a tool it can be readily applied in data analysis processes, such as k-mer clustering, construction of genomic "signatures" from known datasets (e.g. FASTQ reads) etc.

\section*{Conclusion}

This work represents an initial investigative step in applying the formalism of language representation to biological sequences, aiming to identify and quantify the underlying structure. The preliminary results are definitely encouraging and hopefully opening the way to answering more complex questions.

\section*{Acknowledgments}
We would like to acknowledge Kostas Stamatopoulos and Andreas Agathangelidis (\href{http://inab.certh.gr}{INAB|CERTH}) for providing the real-world dataset. 

\section*{Supporting information}



%
%
%

\bibliography{DNA_Coding_And_Godel_Numbering.bib}

\end{document}